# Deep sub-Ångstrom imaging of 2D materials with a high dynamic range detector

Yi Jiang*[1], Zhen Chen*[2], Yimo Han[2], Pratiti Deb[1,2], Hui Gao[3,4], Saien Xie[2,3], Prafull Purohit[1], Mark W. Tate[1], Jiwoong Park[3], Sol M. Gruner[1,5], Veit Elser[1], David A. Muller[2,5]

[1.] Department of Physics, Cornell University, Ithaca, NY 14853, USA

[2.] School of Applied and Engineering Physics, Cornell University, Ithaca, NY 14853, USA

[3.] Department of Chemistry, Institute for Molecular Engineering, and James Franck Institute, University of Chicago, Chicago, IL 60637, USA

[4.] Department of Chemistry and Chemical Biology, Cornell University, Ithaca, NY 14853, USA

[5.] Kavli Institute at Cornell for Nanoscale Science, Ithaca, NY 14853, USA

**ABSTRACT**

**Aberration-corrected optics have made electron microscopy at atomic-resolution a widespread and often essential tool for nanocharacterization. Image resolution is dominated by beam energy and the numerical aperture of the lens ($\alpha$), with state-of-the-art reaching ~0.47 Å at 300 keV. Two-dimensional materials are imaged at lower beam energies to avoid knock-on damage, limiting spatial resolution to ~1 Å. Here, by combining a new electron microscope pixel array detector with the dynamic range to record the complete distribution of transmitted electrons and full-field ptychography to recover phase information from the full phase space, we increased the spatial resolution well beyond the traditional lens limitations. At 80 keV beam energy, our ptychographic reconstructions significantly improved image contrast of single-atom defects in $MoS_2$, reaching an information limit close to $5\alpha$, corresponding to a 0.39 Å Abbe resolution, at the same dose and imaging conditions where conventional imaging modes reach only 0.98 Å.**

**MAIN TEXT**

The ability to image individual atoms is essential for the characterization of structure and defects in two-dimensional (2D) materials[1–3]. In scanning transmission electron microscopy (STEM), the most common technique to achieve atomic resolution is high-angle annular dark-field (ADF) imaging, which records electrons scattered through large angles to form an incoherent image. The maximum spatial information contained in a ADF image (or other incoherent imaging modes) is determined by the momentum transfer across the diameter of the probe forming aperture – i.e. twice the semi-convergence angle($\alpha$)[4,5]. Therefore, obtaining high-resolution images generally requires small wavelength and large apertures, the latter in turn introduces phase-distorting artifacts from geometrical and chromatic aberrations. The demonstration of practical aberration correctors[6,7] has ameliorated this condition significantly, and for the past decade the state-of-the-art for ADF images is about 0.5 Å resolution at 300 keV[8,9], which is sufficient for imaging most bulk materials. On the other hand, the characterization of 2D materials, such as single defect detection and imaging interface or edge structures, always requires lower beam energies (ca. ~20-80 keV) to minimize knock-on damage[2,10,11]. Because lower energies imply longer electron wavelengths, the resolution of ADF imaging is significantly reduced and reaching sub-angstrom resolution is only possible with specialized correctors that correct both geometric and chromatic aberrations or with monochromatic electron beams[12,13]. Moreover, ionization damage, which cannot be avoided by lowering the beam voltage, also restricts the total electron dose applied to the sample, limiting the ultimately achievable signal-to-noise (SNR)[14], further reducing image resolution and contrast.

However, it has long been recognized that the information limit set by diffractive optics is not an ultimate limit[15]. Instead, there is phase information encoded throughout a diffraction pattern formed from a localized electron beam, in form of interference patterns between overlapping scattered beams (Figure 1a), and as the incident localized beam is scanned, the phase information and hence the interference patterns change in a predictable manner that can be used to retrieve the phase differences, an approach known as ptychography[16–18]. While originally conceived to solve the phase problem in crystallography, ptychography has received renewed attention as a dose-efficient technique to recover the projected potential of materials, with modifications to measure finite thickness and three-dimensional samples[19–26]. In principle. the resolution is limited by the largest scattering angle at which meaningful information can still be recorded, however as electron scattering form factors have a very strong angular dependence, the signal falls rapidly with scattering angle, requiring a detector with high dynamic range and sensitivity to exploit this information.

To date, ptychography has been widely adopted for light[27] and x-ray[22,28] applications, yet the technique is still under-explored in transmission electron microscopy in large part because of the detector challenges. Traditional electron cameras such as charge-coupled devices (CCDs) and pixelated detectors have been hampered by slow readout speeds or poor dynamic ranges, previous work[20,29–35] has mainly made use of electrons only within the bright field disk and thus image resolution did not overcome the 2α limit imposed by the physical aperture. The first attempt to demonstrate super-resolution ptychography phased the Fourier coefficients of silicon out to the (400) reflection to reconstruct the unit cell with a resolution of 1.36 Å[17]. However, this result only determined structure factors, limiting its applications to periodic crystalline structures. A more recent demonstration by Humphry *et al.* for a lower-resolution scanning electron microscope equipped with a CCD camera showed that the resolution of iterative ptychographic reconstructions can be improved when utilizing information at higher scattering angles[36].

There are three challenges to improving resolution and dose efficiency to the point needed to advance beyond the current state of the art diffractive imaging. First, a detector must be able to record the full range of scattered intensities without introducing non-linear distortions or saturating the central beam. Second, the detector must not only possess single electron sensitivity, but it must also retain a high detective quantum efficiency (DQE) when summing over the large ranges of empty pixels at high scattering angles. Thirdly, each diffraction pattern must be recorded rapidly enough that the full image is not sensitive to drift and instabilities in the microscope – usually leaving only a few minutes to record a full 4D data set. The combination of the first and third conditions poses an additional constraint that the detector must also have a high dynamic current range – i.e. it is not sufficient to count single electrons for a long time at a low beam current, but rather requires large currents per pixel to be recorded in very short times. Most pulse counting methods are limited to ~2-10 MHz by the transit time of the electron cloud through the silicon detector which translates to 0.3 - 1.6 pA/pixel, although few systems reach this limit, and to keep non-linearities below 10%, a limit of 0.03 pA/pixel is more typical[37]. Direct charge integration in a CCD geometry is even more limited by the well depth, to about 20 electrons/pixel/frame, which at a 1 kHz frame is 0.003 pA/pixel and whose single-frame Poisson statistics would then be below the Rose criterion for contrast detectability[38].

To overcome these challenges, we developed and employed a new type of electron microscope pixel array detector (EMPAD)[39] that is capable of recording all the transmitted electrons with sufficient sensitivity and speed to provide a complete ptychographic reconstruction. Our EMPAD design has a high dynamic range of 1,000,000:1 while preserving single electron sensitivity with a signal-to-noise ratio of 140 for a single electron at 200 keV[39]. The detector retains a good performance from 20-300 keV. Here we operated at 80 keV where the noise per pixel is 1/50 of an electron, the DQE is 0.96, and the maximum beam current per pixel is 5 pA. By utilizing essentially all collected electrons (99.95% of the incident beam), with a full 4D data set acquired in typically a minute, our full-field ptychographic reconstructions roughly double the image resolution compared to the highest-resolution conventional single-channel imaging modes such integrated center-of-mass (iCoM)[40,41] or ADF-STEM.

**Results**

**Data acquisition and reconstruction.** Fig. 1a shows a schematic of the experimental configuration with the EMPAD. To minimize radiation damage, a monolayer of $MoS_2$ is imaged at 80 keV primary beam energy. At each (x, y) scan position, the EMPAD records a diffraction pattern $(k_x, k_y)$ from the convergent probe, thus forming a four-dimensional (4D) data set (x, y, $k_x$, $k_y$). Fig. 1b&c show averaged diffraction patterns corresponding to two positions near a single molybdenum column. Supplementary Video 1 shows a continuous evolution of diffraction patterns at various scan positions, where it is easier to observe the phase changes in the overlaps between higher order diffraction disks. The considerable changes in the distribution outside the central disk provides essential contrast information in ADF images and exploiting the phase information encoded in the contrast between overlapping higher-order disks is responsible for resolution improvement in full-field ptychography over previous bright-field ptychographic methods. The reconstruction algorithm is implemented in house by adopting the extended ptychographic iterative engine (ePIE) algorithm[21,42], which iteratively reconstructs the transmission function and refines the probe function to accommodate aberrations and noise. We also compare our performance to the simpler Wigner distribution deconvolution (WDD)[43], which in its simplest form assumes a known probe function, and shows a similar performance to using ePIE on only the central disk (i.e. with a cutoff $\alpha$), or integrated center-of-mass imaging (iCOM).

The 4D EMPAD data can generate all elastic imaging modes for benchmarking from the same data set, including coherent bright-field (BF), iCOM and ADF. As shown in Fig. 2a&e, the coherent BF image has the poorest resolution (restricted to within $\alpha$ as expected) and atoms are barely visible. The incoherent ADF image (Fig. 2b&f) "doubles" the information limit (i.e. from $\alpha$ to $2\alpha$) but is limited by low signal-to-noise ratio and residual probe aberrations. Although the iCoM image is less noisy, its resolution is still within $2\alpha$ (Fig. 2g) as the structural information is influenced by the incident probe via convolution. In contrast, full-field ptychography directly recovered the phase of the transmission function (Fig. 2d) and achieved an information limit of $5\alpha$ (Fig. 2h). Noise artifacts are also reduced significantly, and the light-atom sulfur monovacancy (indicated by red arrows) is more clearly resolved. Fig. 2i&j shows an enlarged section of the Fourier intensity map from the ptychographic reconstruction and a line profile across a diffraction spot at the $5\alpha$ limit, demonstrating an estimated Abbe resolution[44] of 0.39 Å or better (there are still higher-order spots of weaker intensity but not as uniform in all directions). For comparison, with our electron optical conditions, the expected Abbe resolution for conventional incoherent imaging modes such as ADF-STEM would be $2\alpha$ or 0.98 Å.

A second measure of spatial resolution is that of a minimum resolvable distance between two atoms. For 2D materials this is complicated by the fact that this would require atoms spaced closer than the shortest known bond lengths. To accomplish this test, we use a twisted bilayer sample of two $MoS_2$ sheets rotated by 6.8° to each other. This is effectively an atomic moiré pattern that has the effect of providing projected Mo-Mo atomic distances that vary from a full bond length apart to fully superimposed atoms, with many intermediate distances across the moiré quasiperiodicity of 28 Å – e.g. Fig 1c of van der Zande et al[45]. Figure 3 shows the ptychographic reconstruction across a moiré supercell, in which atomic columns midway between the aligned regions are resolved as separate atoms at 0.85±0.02 Å. The dip between adjacent columns can still be seen at 0.60±0.02 Å – close to the Raleigh limit for resolution. Atom pair peaks measured at 0.42±0.02 Å show a 6% dip at the midpoint. From a rigid model structure of the rotated bilayer, assuming no relaxation occurs (and some probably does), the model separations for these atom pairs marked on Figure 3 would be 0.87 Å, 0.60 Å, and 0.36 Å respectively. While not all atoms can be reconstructed because of scan noise, we have multiple moiré repeats to distinguish random from systematic errors. Ignoring source size contributions, the expected Raleigh limit for an incoherent imaging mode such as ADF imaging for this experimental condition would be 1.2 Å. In other words, our full field ptychographic reconstruction demonstrates double the Raleigh resolution compared to conventional 2α imaging methods. Moreover, some closely-spaced atoms lose the central dip at just below 0.40 Å, which would be the Sparrow criterion for resolution, close to the Abbe limit estimated from Figure 2.

To understand how dark-field electrons contribute to resolution improvement, we performed additional reconstructions using diffraction patterns with outer cutoff angles varying from one to four times the aperture size. As shown in Fig. 4a-h, when only using the central bright-field disk, the reconstructed phase has a relatively low resolution, similar to that of the ADF and iCoM images. As the cutoff increases, atoms become sharper and more clearly resolved. Beyond 3α, where there are fewer scattered electrons, the improvements become less obvious and the reconstruction is mainly limited by the total electron dose. As discussed in more detail in the following section, increasing the collection angle beyond where there is meaningful signal in the diffraction pattern does not introduce high-spatial-frequency artifacts. Instead, the reconstruction retains its limiting form. The same trend can be seen in simulations (2$^{nd}$ column of Supplementary Fig. 1).

As a test of linearity, Fig. 4i shows that the phase at the sulfur monovacancy position, where only one sulfur atom is present, is about half of the phase shift of the two-sulfur sites, validating the strong phase approximation and ePIE reconstruction for these thin 2D materials. That the reconstructed probes (Fig. 4j&k) have similar shapes at different cutoffs also indicates that it is the dark-field electrons that contribute to resolution improvement.

**Influence of total electron dose**. We further explored the potential limits of full-field ptychography using simulated data sets for a wide range of collection angles and beam currents, including cases when the cutoff is extended beyond most of the scattered electrons, as well as when the total dose is too small for a stable reconstruction to be achieved. The image quality is evaluated by both the range of the reconstructed phase and the root mean squared width of the molybdenum atoms measured from the standard deviation of a Gaussian fit. At high dose, one would expect the ultimate information limit of the ptychographic reconstruction to be twice the cutoff angle – i.e. 8α for a 4α cutoff. In practice, as shown in Fig. 5, ptychographic reconstructions are mainly influenced by the total electron dose and these limits are not reached for the larger cutoffs. There is only a slight improvement between 3α and 4α cutoff at a typical operating beam current (1-50 pA), which agrees with experimental data in Fig. 4. If the beam current is too low (e.g. 0.01 pA corresponding to a dose of 260 e$^-$/Å$^2$), atoms become distorted with reduced image resolution, yet some of the overall structure of $MoS_2$ is still recognizable (Fig. 5c). As the beam current increases, the influence of Poisson noise becomes less significant as there are sufficient electrons scattered into high angles to provide interference between high indexed lattice planes in the crystalline structure. At higher doses, the resolution of ptychographic reconstruction is able to benefit more fully from the increased maximum collection angle. Because the EMPAD has a high DQE, increasing the cutoff angle beyond where there is signal does not compromise resolution or introduce additional artifacts, as shown by the fact all curves in Fig. 5a collapse to the same trend as beam currents decreases.

Fig. 6 compares the performance of ePIE and WDD ptychographic reconstructions at low electron dose. Using the same datasets simulated with a small in-focused probe, both methods show similar results and can achieve atomic resolution at ~500 e⁻/Å². On the other hand, using a large defocused probe, the ePIE technique can further improve reconstruction quality beyond that of WDD (Fig. 6c). Overall, ptychographic reconstructions are seen to be more dose-efficient than low-angle ADF (integrating from 1α upwards) (Fig. 6d), the most dose-efficient single-channel STEM imaging mode for single-atom detection to date[46]. For the more typical range of ADF angles such as the experimental data of Fig. 2b, the effects are even pronounced (e.g. Fig. 2d). As shown in Supplementary Fig. 2, the advantage of ptychography over ADF becomes more significant for materials with lighter elements such as graphene.

**Discussion**

In addition to beam current and detector configuration, other practical sources of errors such as sample contamination and scanning drift may cause distortions and reduce reconstruction quality. However, to date we have found full-field ptychography can outperform all other techniques we have tested under same conditions (e.g. Supplementary Fig. 4). By incorporating other physical constraints and prior knowledge, we envisage more advanced reconstruction strategies, when applied to full-field electron ptychography data, can compensate for inaccurate scan positions or make allowance for thick specimens with strong dynamic scattering.

In summary, we demonstrated that with the entire distribution of scattered electrons collected by the EMPAD, full-field ptychography significantly enhances image resolution and contrast compared to traditional electron imaging techniques, even at low beam voltages. With the improved detector, atomic-scale ptychographic reconstructions are no longer restricted by the aperture size. Instead, image quality is determined by the electron dose and collection angle. The technique provides an efficient tool for unveiling sub-angstrom features of 2D or dose-sensitive materials. Combined with the emerging ultra-low voltage aberration-corrected microscopes, the technique has the potential to tackle currently hard problems such as direct imaging of lattice displacements in twisted-layer structures, structural distortions around single atom dopants and vacancies, and even 3D tomography.

**Methods**

**EMPAD data acquisition.** The 4D dataset of monolayer $MoS_2$ was taken using an aberration-corrected FEI Titan with 9.5 pA beam current, 80 keV beam energy and 21.4 mrad aperture size, with the dose limited by the radiation resistance of the sample. The EMPAD has 128 x 128 pixels and a readout speed of 0.86 ms/frame. 51 x 87 diffraction patterns with scan step size of 0.19 Å are used to generate BF, ADF, and iCoM images in Fig. 2a-c. The ADF image is integrated from 64.2 mrad ($3\alpha$) to 84.6 mrad ($4\alpha$). Higher angles did not add significant contributions to the signal. The dataset of twisted $MoS_2$ in Fig. 3 was taken with the same conditions except for a 11.7 pA beam current and 0.59 Å scan step size.

**Ptychographic reconstructions.** Prior to reconstruction, all diffraction patterns are padded with zeros to a total size of 256 x 256 and thus the pixel size in reconstructed phase is 0.12 Å/pixel. The ePIE method[21] is implemented in-house with modifications to exclude bad pixels in the diffraction patterns. The algorithm aims to minimize the Euclidian distance between reconstructed and measured diffraction patterns. In general, the convergence of reconstructions depends on the number of iterations and update parameters for the transmission function and the probe function. Because experimental data contains noise and other sources of errors, fast convergence may introduce noisy artifacts and reduce reconstruction quality[47,48]. To alleviate this problem, we used a small update parameter (0.1) for the transmission function and the reconstruction of the probe function was limited to data taken in areas with minimal contamination.

**Fourier resolution estimation.** All diffractograms (Fourier intensity) of BF, ADF, iCoM, and ptychography reconstructions in Fig. 2 are calculated based on images constructed from 128 x 128 diffraction patterns (see Supplementary Fig. 4). To visualize diffraction spots, a periodic and smooth decomposition[49] is applied to images to reduce artifacts caused by edge discontinuities, followed by a light Gaussian filter, making the diffraction spots slightly larger and thus more visible. The Fourier intensity was rescaled to enhance intensity of higher order spots for better visualization.

**4D data simulation.** All datasets used for dose-cutoff simulations were generated by the μSTEM software. For the in-focused probe, 21 x 24 diffraction patterns with 0.45 Å scan step size is simulated at 80 keV beam energy and with 21.4 mrad aperture size. Thermal diffuse scattering effect is included with the frozen phonon approximation. The diffraction patterns are further corrupted with Poisson noise determined by the simulated beam dose. Supplementary Fig. 1 shows selected ePIE reconstructions at different beam currents (from 0.01 pA to 100 pA) and cutoff angles ($1\alpha$-$4\alpha$). Simulations (Fig. 6c) of large probe profile used 80nm defocus and 10.04 Å scan step size.

**Data availability.** All relevant data are available from the corresponding author (david.a.muller@cornell.edu) upon request.

**FIGURES**

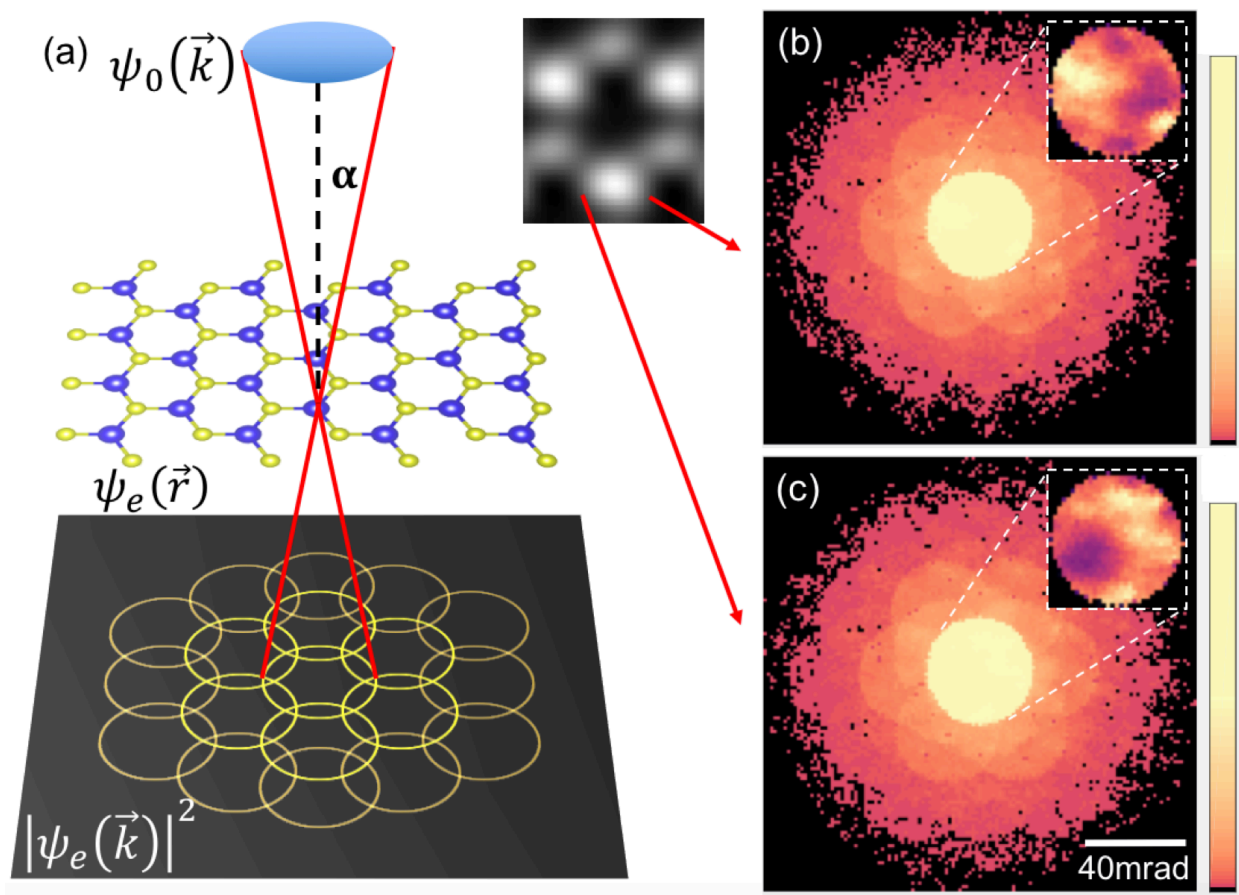

**Figure 1. Scanning transmission electron microscopy imaging using the electron microscope pixel array detector.** (a) At each scan position, the incident probe ($\psi_0(\vec{k})$) is focused on the sample and the entire diffraction pattern of the exit wave ($|\psi_e(\vec{k})|^2$) is recorded by the EMPAD. (b&c) Averaged diffraction patterns (on a log-scale) from electron beam at the marked scan positions near a molybdenum column with insets showing the intensity (on a linear-scale) of the bright field disks. The significant intensity differences at large scattering angles provide contrast information for ADF imaging and are essential for resolution enhancement in ptychography.

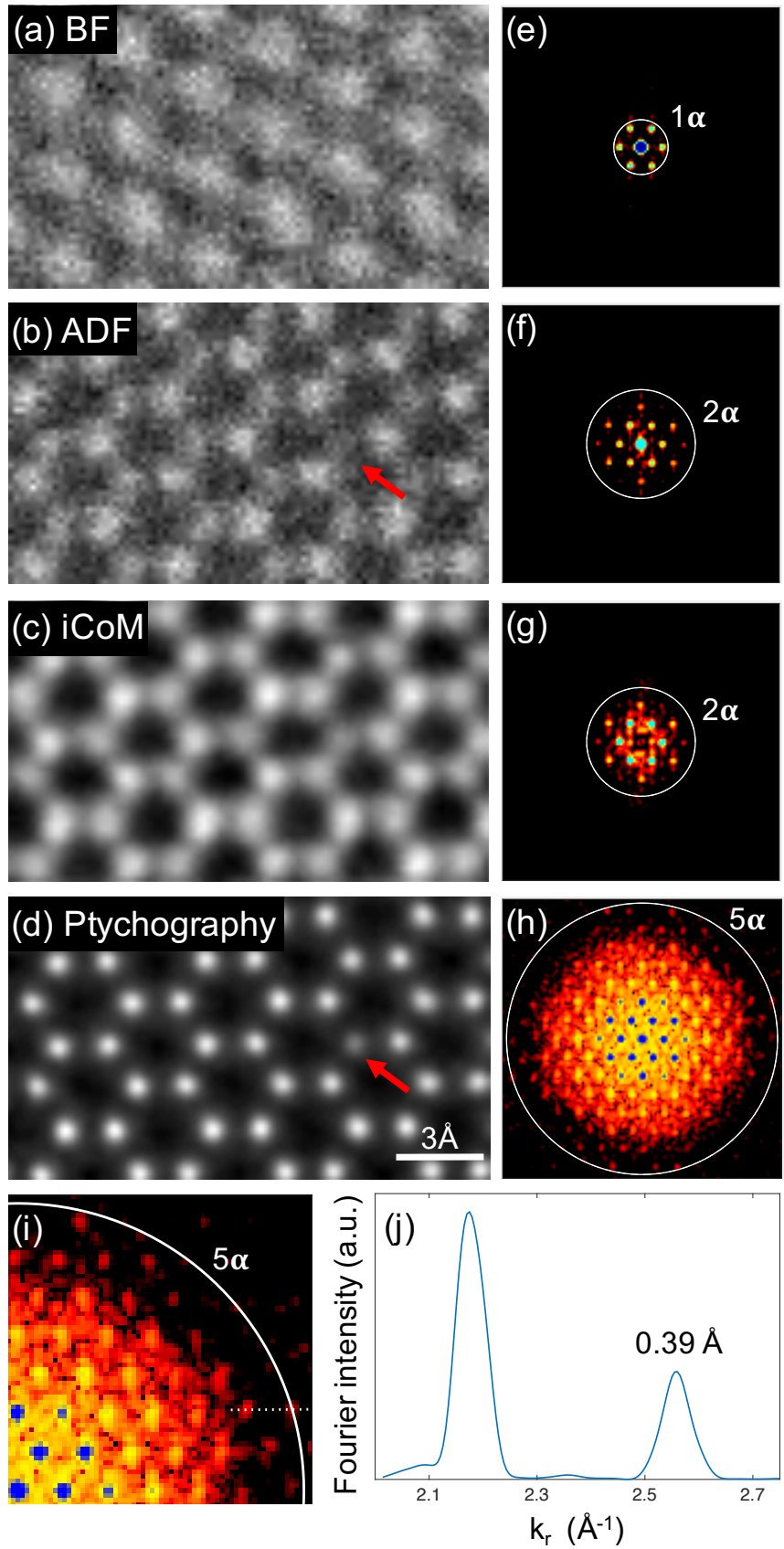

**Figure 2. Comparison of different imaging techniques using the same measured 4D EMPAD dataset from a monolayer of MoS$_2$.** (a) Coherent bright field (BF) image. (b) Incoherent annular dark field (ADF) image. (c) Integrated center-of-mass image. (d) Phase of transmission function reconstructed by full-field ptychography. The red arrows indicate a sulfur monovacancy that is readily detectable in ptychograhy. (e-h) False-color diffractograms (on log-scale) of BF, ADF, iCoM and full-field ptychography reconstruction, respectively. The information limit of ptychography is close to 5α (107 mrad). (i) Zoomed-in diffractogram (on log scale) of Figure h. (j) A line profile (on linear scale) across two diffraction spots. The peak at five times the aperture size corresponds to an Abbe resolution of 0.39 Å.

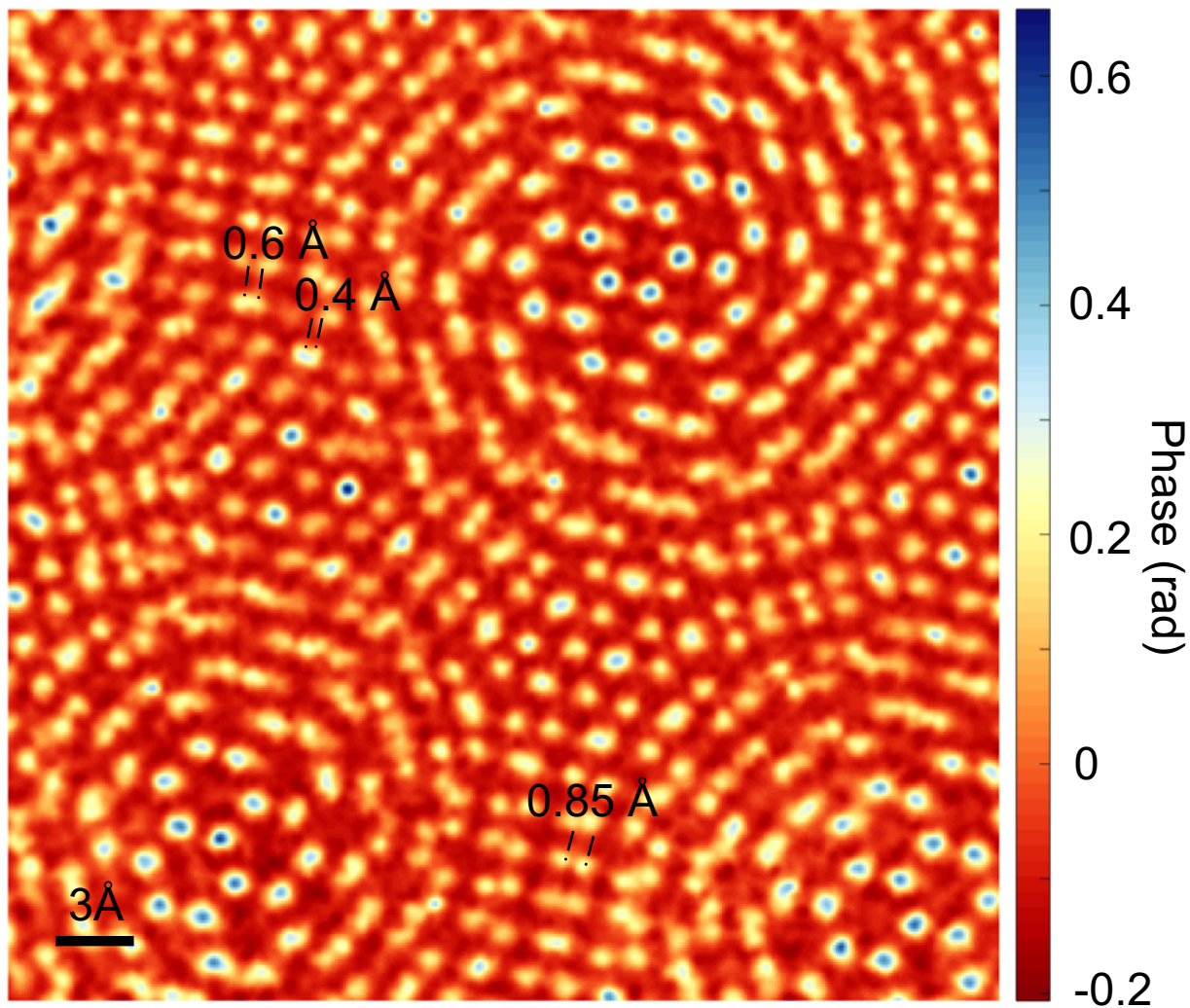

**Figure 3. Real-space resolution test of full field ptychography using twisted bilayer MoS₂.** The two sheets are rotated 6.8° degrees to each other, and the misregistration of the Mo atoms provides a range of projected distances varying from a full bond length down to complete overlap. Atoms are still cleanly resolved at 0.85±0.02 Å separation, with a small dip still present at about 0.61±0.02 Å that would correspond to similar contrast expected for the Rayleigh criterion for conventional imaging. Atom pair peaks at 0.42±0.02 Å show a 6% dip at the midpoint, suggesting the Sparrow limit should lie just below 0.4 Å. The diffraction-limited Raleigh criterion for direct imaging methods such as ADF-STEM would be 1.2 Å for our electron optical configuration. Line profiles through atoms pairs can be found in supplementary Figure 3.

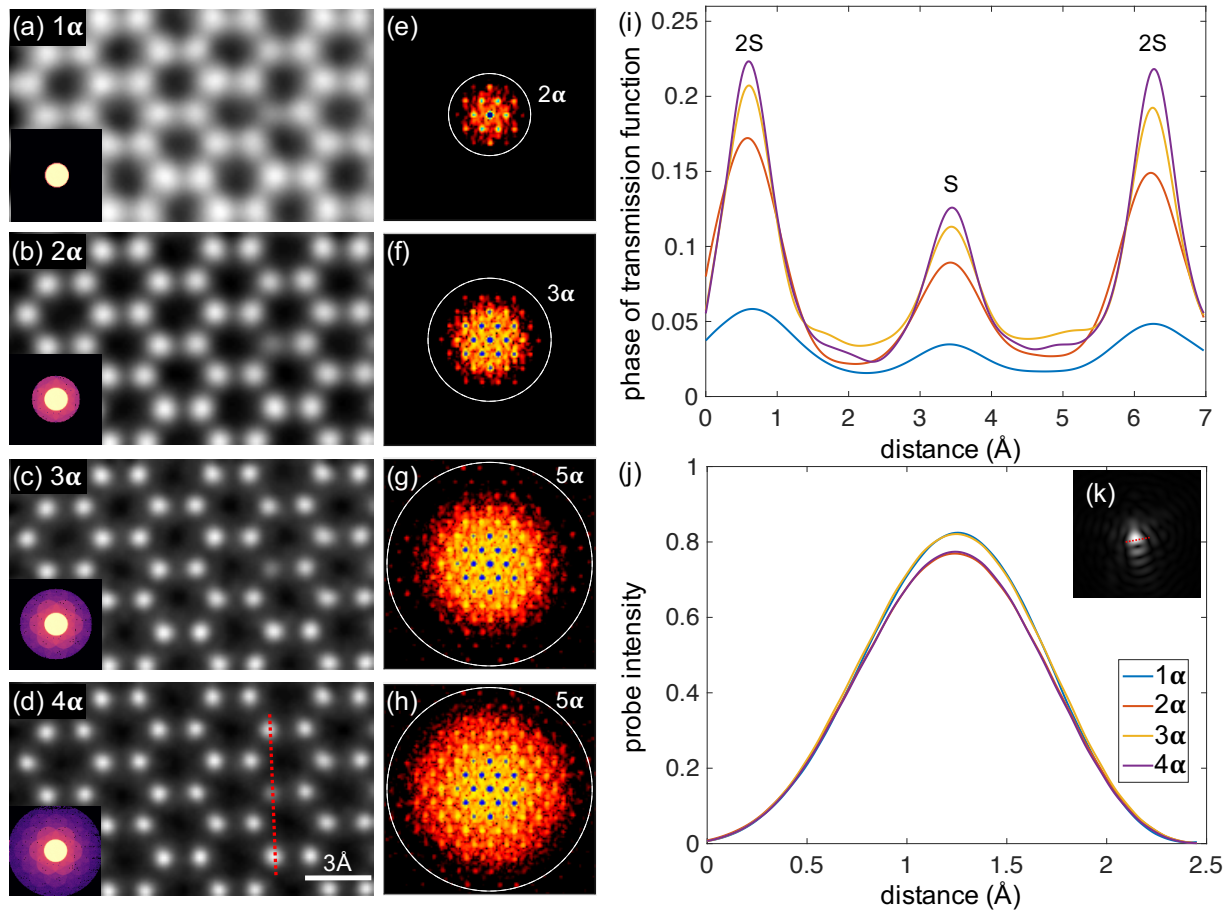

**Figure 4. Ptychographic reconstructions using data with different cutoff angles.** (a-d) Ptychographic reconstructions using electrons collected with cutoffs from 1-4 times the aperture size (α). The averaged diffraction patterns are shown in the lower-left corner and false-color diffractograms (on log-scale) of the reconstructions are shown in (e-h), respectively. (i) Line profiles across three Sulfur columns, as indicated by the dashed line in (d). (j) Line profiles across the reconstructed probe function at different cutoffs. (k) Probe profile reconstructed by ptychography using the data set with a 4α cutoff.

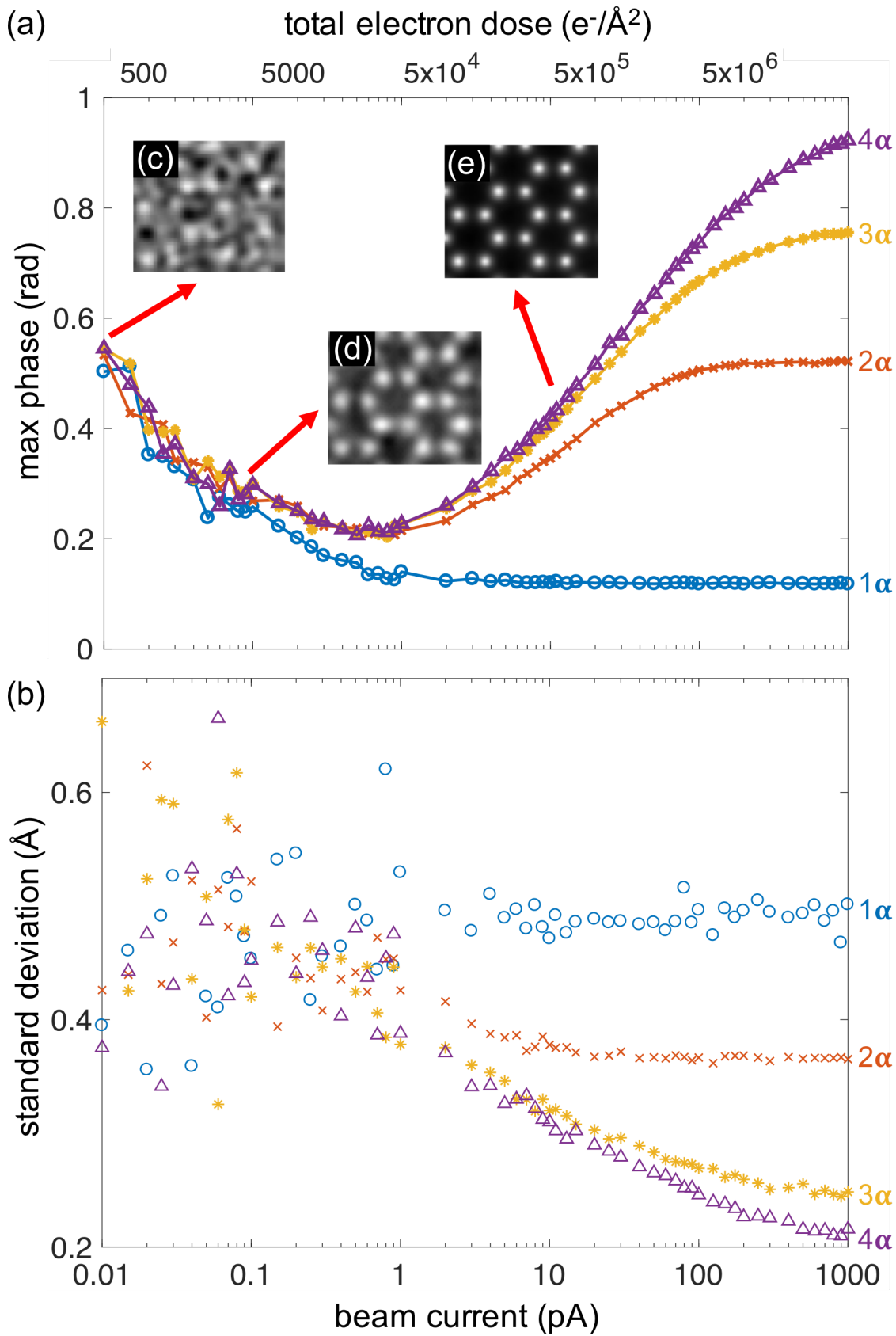

**Figure 5. Simulation study of full-field ptychography as a function of cutoff angle and beam current.** The resolution of reconstruction is evaluated by (a) the maximum range of the reconstructed phase and (b) Fitted Gaussian standard deviation of molybdenum atoms. At large electron doses, the resolution is determined by the maximum detector angle. As current decreases, the resolution is instead limited by the Poisson noise. (c-e) Reconstructed phase maps using diffraction patterns with $4\alpha$ cutoff at 0.01 pA, 0.1 pA and 10 pA, respectively.

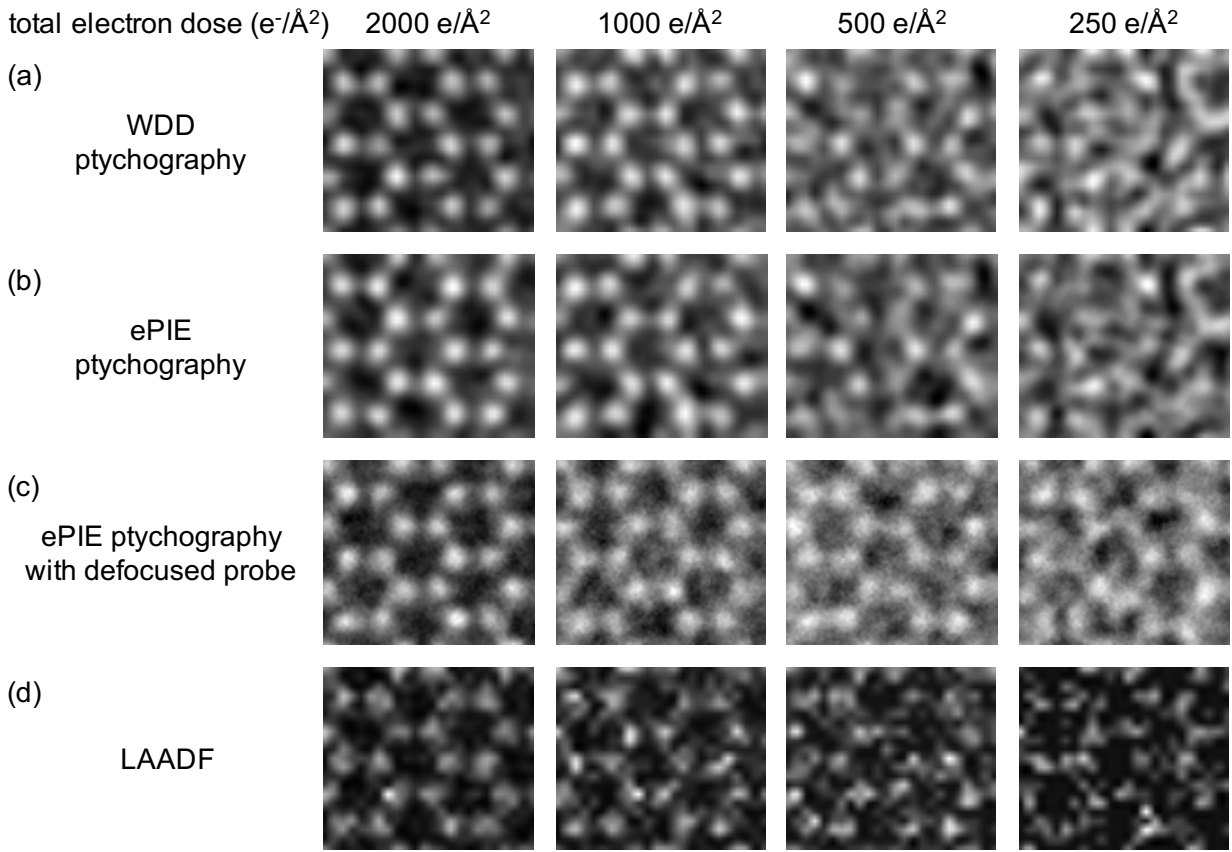

**Figure 6. Comparison between ptychographic techniques and low-angle ADF imaging at low electron dose.** (a&b) Ptychographic reconstructions of simulated data with an in-focused probe using the WDD and ePIE methods, respectively. (c) ePIE reconstructions of data with a large defocused probe. (d) Low-angle ADF (integrating from $1\alpha$ to $4\alpha$) using the same simulated dataset as (a&b). The defocused ePIE approach shows a roughly factor of two or better advantage over the other two ptychography approaches, and a roughly fourfold advantage over LAADF imaging, the optimal single-channel imaging method for 2D materials to date.

**ASSOCIATED CONTENT**

**Author Contributions** Y.J. and Z.C. contributed equally to this work. Experiments were performed and designed by Z. C., Y. H., and D.A.M. Y.J. contributed to data analysis and ptychographic reconstruction with support from V.E. Sample preparation by P. D. and Y. H., from $MoS_2$ thin films synthesized by H.G, S.X. and J.P. EMPAD optimization by P.P., M.W.T. and S.M.G. All authors discussed the results and implications throughout the investigation. All authors have given approval to the final version of the manuscript.

**ACKNOWLEDGMENTS** Y.J. and V.E. acknowledges support from DOE grant DE-SC0005827. Z.C and D.A.M. are supported by the PARADIM Materials Innovation Platform program in-house program by NSF grant DMR-1539918. Electron microscopy facility and support for Y.H. from the NSF MRSEC program (DMR 1719875) and NSF MRI grant DMR-1429155. H.G., S.X, and J.P. acknowledge additional support from AFOSR MURI (FA9550-16-1-003) and UChicago NSF MRSEC program (DMR 1420709). Detector development at Cornell by S.M.G. supported by the DOE (DE-SC0017631). We thank Kayla Nguyen, Pinshane Huang, Martin Humphry and Peter Nellist for useful discussions, and Bin Jiang from Thermo Scientific for help during the initial experiments.

**AUTHOR INFORMATION** Reprints and permissions information is available at www.nature.com/reprints. The authors note a potential competing financial interest in that Cornell University has licensed the EMPAD hardware to Thermo Scientific. Readers are welcome to comment on the online version of this article at www.nature.com/nature.

Correspondence and requests for materials should be addressed to D.A.M. (david.a.muller@cornell.edu).

**REFERENCES**

1.  Meyer, J. C., Girit, C. O., Crommie, M. F. & Zettl, A. Imaging and dynamics of light atoms and molecules on graphene. *Nature* **454,** 319–322 (2008).

2.  Krivanek, O. L. *et al.* Atom-by-atom structural and chemical analysis by annular dark-field electron microscopy. *Nature* **464,** 571–574 (2010).

3.  Huang, P. Y. *et al.* Grains and grain boundaries in single-layer graphene atomic patchwork quilts. *Nature* **469,** 389–392 (2011).

4.  Sparrow, C. M. On Spectroscopic Resolving Power. *Astrophys. J.* **44,** 76 (1916).

5.  Black, G. & Linfoot, E. H. Spherical aberration and the information content of optical images. *Proc R Soc Lond A* **239,** 522–540 (1957).

6.  Haider, M. *et al.* Electron microscopy image enhanced. *Nature* **392,** 768–769 (1998).

7.  Batson, P. E., Dellby, N. & Krivanek, O. L. Sub-ångstrom resolution using aberration corrected electron optics. *Nature* **418,** 617–620 (2002).

8.  Erni, R., Rossell, M. D., Kisielowski, C. & Dahmen, U. Atomic-Resolution Imaging with a Sub-50-pm Electron Probe. *Phys. Rev. Lett.* **102,** 096101 (2009).

9.  Sawada, H. *et al.* STEM imaging of 47-pm-separated atomic columns by a spherical aberration-corrected electron microscope with a 300-kV cold field emission gun. *J. Electron Microsc. (Tokyo)* **58,** 357–361 (2009).

10. Kaiser, U. *et al.* Transmission electron microscopy at 20kV for imaging and spectroscopy. *Ultramicroscopy* **111,** 1239–1246 (2011).

11. Meyer, J. C. *et al.* Accurate Measurement of Electron Beam Induced Displacement Cross Sections for Single-Layer Graphene. *Phys. Rev. Lett.* **108,** 196102 (2012).

12. Sawada, H., Sasaki, T., Hosokawa, F. & Suenaga, K. Atomic-Resolution STEM Imaging of Graphene at Low Voltage of 30 kV with Resolution Enhancement by Using Large Convergence Angle. *Phys. Rev. Lett.* **114,** 166102 (2015).

13. Linck, M. *et al.* Chromatic Aberration Correction for Atomic Resolution TEM Imaging from 20 to 80 kV. *Phys. Rev. Lett.* **117,** 076101 (2016).

14. Henderson, R. The potential and limitations of neutrons, electrons and X-rays for atomic resolution microscopy of unstained biological molecules. *Q. Rev. Biophys.* **28,** 171–193 (1995).

15. Gabor, D. A New Microscopic Principle. *Nature* **161,** 777–778 (1948).

16. Hoppe, W. Beugung im inhomogenen Primärstrahlwellenfeld. I. Prinzip einer Phasenmessung von Elektronenbeugungsinterferenzen. *Acta Crystallogr. A* **25,** 495–501 (1969).

17. Nellist, P. D., McCallum, B. C. & Rodenburg, J. M. Resolution beyond the 'information limit' in transmission electron microscopy. *Nature* **374,** 630–632 (1995).

18. Nellist, P. D. & Rodenburg, J. M. Electron Ptychography. I. Experimental Demonstration Beyond the Conventional Resolution Limits. *Acta Crystallogr. A* **54,** 49–60 (1998).

19. Li, P., Edo, T. B. & Rodenburg, J. M. Ptychographic inversion via Wigner distribution deconvolution: Noise suppression and probe design. *Ultramicroscopy* **147,** 106–113 (2014).

20. Yang, H. *et al.* Simultaneous atomic-resolution electron ptychography and Z-contrast imaging of light and heavy elements in complex nanostructures. *Nat. Commun.* **7,** 12532 (2016).

21. Maiden, A. M. & Rodenburg, J. M. An improved ptychographical phase retrieval algorithm for diffractive imaging. *Ultramicroscopy* **109,** 1256–1262 (2009).

22. Thibault, P. *et al.* High-Resolution Scanning X-ray Diffraction Microscopy. *Science* **321,** 379–382 (2008).

23. Pelz, P. M., Qiu, W. X., Bücker, R., Kassier, G. & Miller, R. J. D. Low-dose cryo electron ptychography via non-convex Bayesian optimization. *Sci. Rep.* **7,** 9883 (2017).

24. Maiden, A. M., Humphry, M. J. & Rodenburg, J. M. Ptychographic transmission microscopy in three dimensions using a multi-slice approach. *JOSA A* **29,** 1606–1614 (2012).

25. Thibault, P. & Menzel, A. Reconstructing state mixtures from diffraction measurements. *Nature* **494,** 68–71 (2013).

26. Maiden, A. M., Sarahan, M. C., Stagg, M. D., Schramm, S. M. & Humphry, M. J. Quantitative electron phase imaging with high sensitivity and an unlimited field of view. *Sci. Rep.* **5,** srep14690 (2015).

27. Rodenburg, J. M., Hurst, A. C. & Cullis, A. G. Transmission microscopy without lenses for objects of unlimited size. *Ultramicroscopy* **107,** 227–231 (2007).

28. Rodenburg, J. M. *et al.* Hard-X-Ray Lensless Imaging of Extended Objects. *Phys. Rev. Lett.* **98,** 034801 (2007).

29. Hüe, F., Rodenburg, J. M., Maiden, A. M., Sweeney, F. & Midgley, P. A. Wave-front phase retrieval in transmission electron microscopy via ptychography. *Phys. Rev. B* **82,** 121415 (2010).

30. Hüe, F., Rodenburg, J. M., Maiden, A. M. & Midgley, P. A. Extended ptychography in the transmission electron microscope: Possibilities and limitations. *Ultramicroscopy* **111,** 1117–1123 (2011).

31. Putkunz, C. T. *et al.* Atom-Scale Ptychographic Electron Diffractive Imaging of Boron Nitride Cones. *Phys. Rev. Lett.* **108,** 073901 (2012).

32. D'Alfonso, A. J. *et al.* Deterministic electron ptychography at atomic resolution. *Phys. Rev. B* **89,** 064101 (2014).

33. Pennycook, T. J. *et al.* Efficient phase contrast imaging in STEM using a pixelated detector. Part 1: Experimental demonstration at atomic resolution. *Ultramicroscopy* **151,** 160–167 (2015).

34. Yang, H. *et al.* Electron ptychographic phase imaging of light elements in crystalline materials using Wigner distribution deconvolution. *Ultramicroscopy* **180,** 173–179 (2017).

35. Wang, P., Zhang, F., Gao, S., Zhang, M. & Kirkland, A. I. Electron Ptychographic Diffractive Imaging of Boron Atoms in LaB 6 Crystals. *Sci. Rep.* **7,** 2857 (2017).

36. Humphry, M. J., Kraus, B., Hurst, A. C., Maiden, A. M. & Rodenburg, J. M. Ptychographic electron microscopy using high-angle dark-field scattering for sub-nanometre resolution imaging. *Nat. Commun.* **3,** 730 (2012).

37. Frojdh, E. *et al.* Count rate linearity and spectral response of the Medipix3RX chip coupled to a 300μm silicon sensor under high flux conditions. *J. Instrum.* **9,** C04028 (2014).

38. Rose, A. *Vision Human And Electronic*. (Plenum Press, 1949).

39. Tate, M. W. *et al.* High Dynamic Range Pixel Array Detector for Scanning Transmission Electron Microscopy. *Microsc. Microanal.* **22,** 237–249 (2016).

40. Close, R., Chen, Z., Shibata, N. & Findlay, S. D. Towards quantitative, atomic-resolution reconstruction of the electrostatic potential via differential phase contrast using electrons. *Ultramicroscopy* **159, Part 1,** 124–137 (2015).

41. Lazić, I., Bosch, E. G. T. & Lazar, S. Phase contrast STEM for thin samples: Integrated differential phase contrast. *Ultramicroscopy* **160,** 265–280 (2016).

42. Maiden, A. M., Humphry, M. J., Zhang, F. & Rodenburg, J. M. Superresolution imaging via ptychography. *JOSA A* **28,** 604–612 (2011).

43. Rodenburg, J. M. & Bates, R. H. T. The Theory of Super-Resolution Electron Microscopy Via Wigner-Distribution Deconvolution. *Philos. Trans. R. Soc. Lond. Math. Phys. Eng. Sci.* **339,** 521–553 (1992).

44. Abbe, E. The Relation of Aperture and Power in the Microscope. *J. R. Microsc. Soc.* **2,** 300–309 (1882).

45. van der Zande, A. M. *et al.* Tailoring the Electronic Structure in Bilayer Molybdenum Disulfide via Interlayer Twist. *Nano Lett.* **14,** 3869–3875 (2014).

46. Hovden, R. & Muller, D. A. Efficient elastic imaging of single atoms on ultrathin supports in a scanning transmission electron microscope. *Ultramicroscopy* **123,** 59–65 (2012).

47. Zuo, C., Sun, J. & Chen, Q. Adaptive step-size strategy for noise-robust Fourier ptychographic microscopy. *Opt. Express* **24,** 20724–20744 (2016).

48. Maiden, A., Johnson, D. & Li, P. Further improvements to the ptychographical iterative engine. *Optica* **4,** 736–745 (2017).

49. Hovden, R., Jiang, Y., Xin, H. L. & Kourkoutis, L. F. Periodic Artifact Reduction in Fourier Transforms of Full Field Atomic Resolution Images. *Microsc. Microanal.* **21,** 436–441 (2015).

Supporting Information For:

# Deep sub-Ångstrom imaging of 2D materials with a high dynamic range detector

*Yi Jiang\*[1], Zhen Chen\*[2], Yimo Han[2], Pratiti Deb[1,2], Hui Gao[3,4], Saien Xie[2,3], Prafull Purohit[1],*

*Mark W. Tate[1], Jiwoong Park[3], Sol M. Gruner[1,5], Veit Elser[1], David A. Muller[2,5]*

[1.] Department of Physics, Cornell University, Ithaca, NY 14853, USA

[2.] School of Applied and Engineering Physics, Cornell University, Ithaca, NY 14853, USA

[3.] Department of Chemistry, Institute for Molecular Engineering, and James Franck Institute, University of Chicago, Chicago, IL 60637, USA

[4.] Department of Chemistry and Chemical Biology, Cornell University, Ithaca, NY 14853, USA

[5.] Kavli Institute at Cornell for Nanoscale Science, Ithaca, NY 14853, USA

**SUPPLEMENTARY FIGURES**

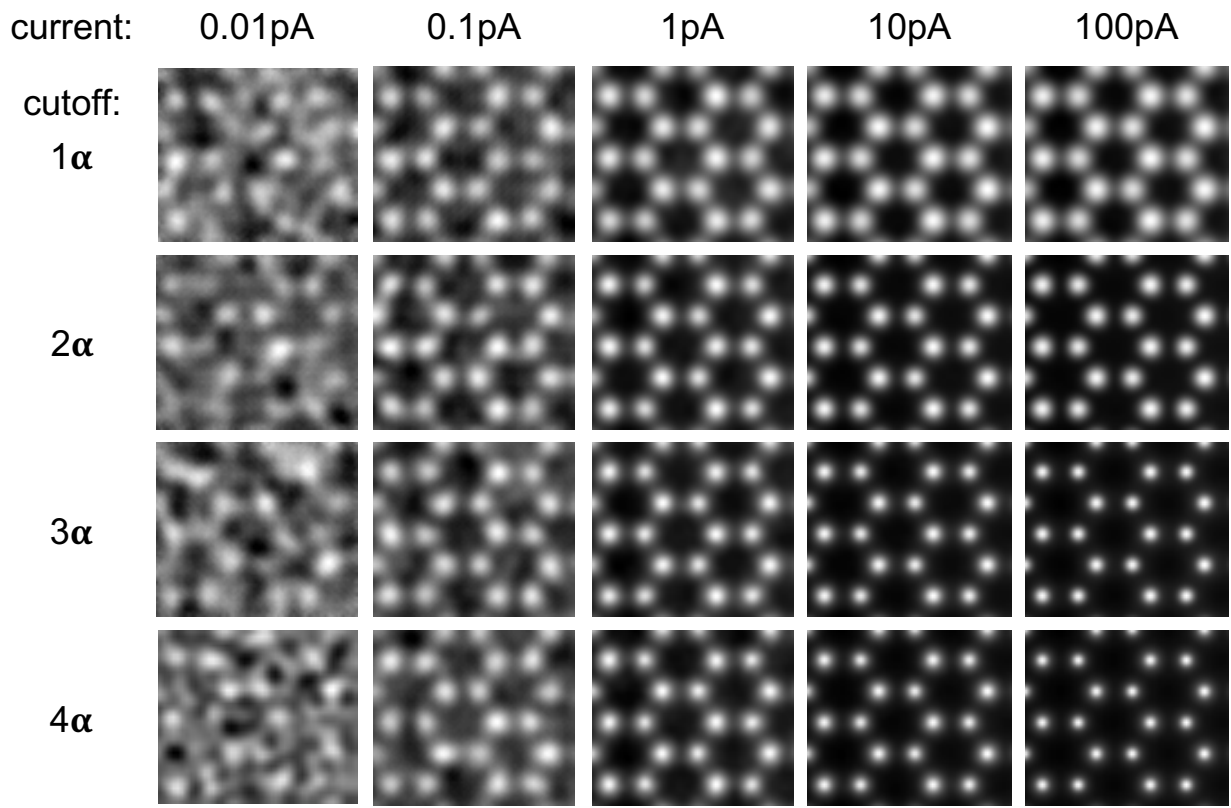

**Supplementary Figure 1. Ptychographic reconstructions of monolayer MoS$_2$ using simulated diffraction patterns at different beam currents and cutoff angles.** At high beam current, the resolution of ptychography reconstruction is fundamentally determined by detector's collection angle. As the beam current decreases, the resolution become dose-limited and distorting artifacts start to appear in the ePIE reconstruction. Beam energy is 80 keV and aperture size (α) is 21.4 mrad.

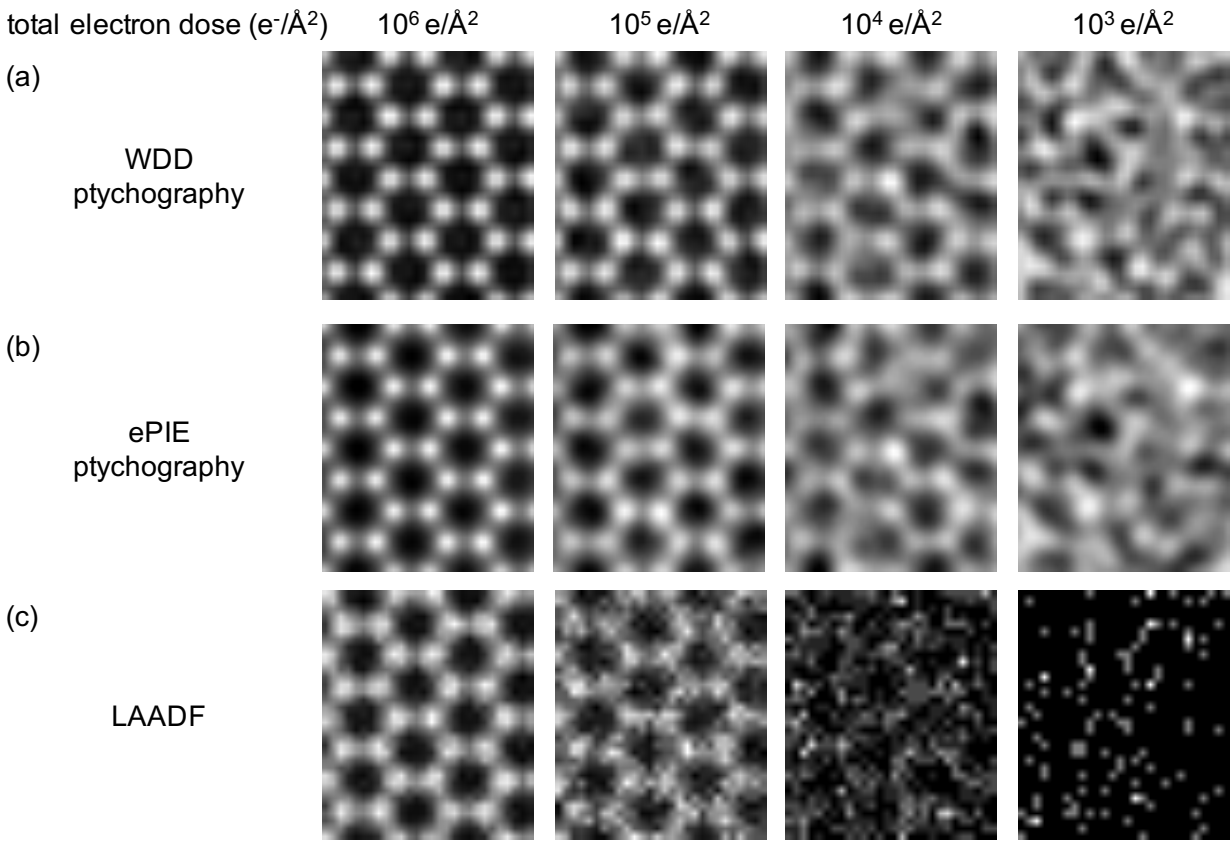

total electron dose (e⁻/Å²) — $10^6$ e/Å² — $10^5$ e/Å² — $10^4$ e/Å² — $10^3$ e/Å²

(a) WDD ptychography

(b) ePIE ptychography

(c) LAADF

**Supplementary Figure 2. Comparison between ptychography techniques and low-angle ADF imaging of graphene.** (a&b) Ptychographic reconstructions of simulated data with an in-focused probe using the WDD and ePIE methods, respectively. (c) Low-angle ADF (integrating from $1\alpha$ to $4\alpha$) using the same simulated dataset as (a&b). Both ptychographic methods show similar reconstructions and are ~10 times more dose-efficient than low-angle ADF technique. Beam energy is 80 keV and aperture size ($\alpha$) is 21.4 mrad.

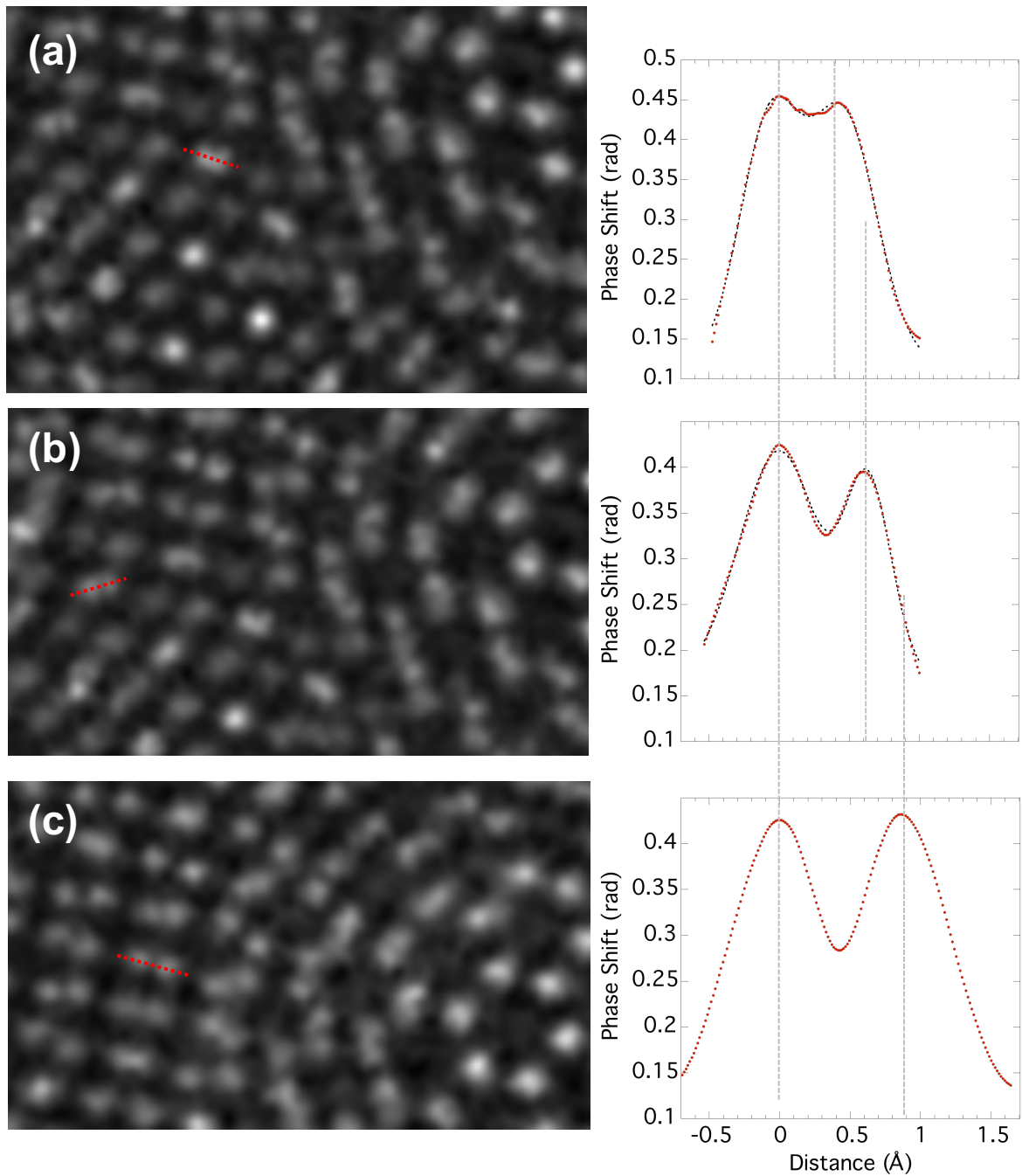

**Supplementary Figure 3. Line profiles through Atom Pairs in the Twisted Bilayer MoS₂ of Figure 3 for measured peak separations of** (a) 0.42±0.02 Å (b) 0.61±0.02 Å and (c) 0.85±0.02 Å. In the line profiles, the black lines are the original intensity and the red lines are the fitted intensity with two Gaussian functions.

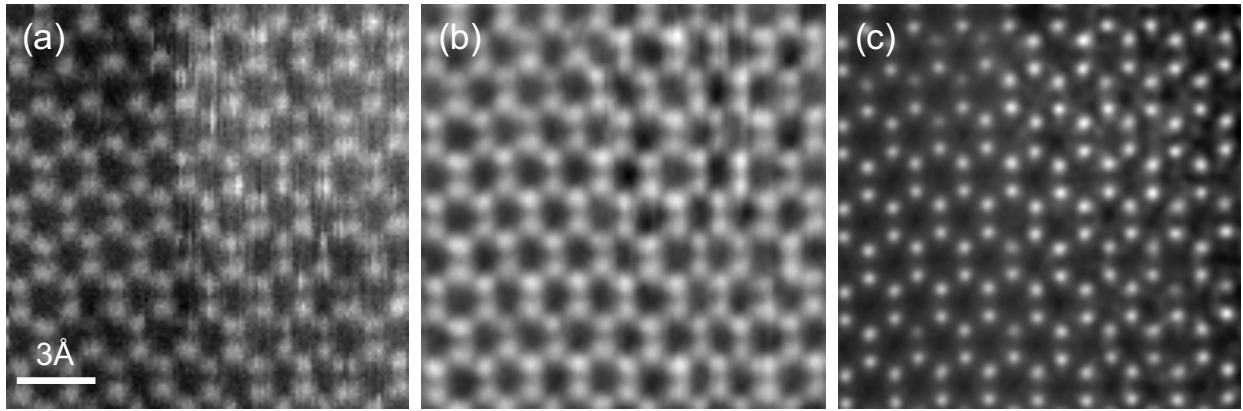

**Supplementary Figure 4. Influence of scanning drift and contamination.** (a) ADF (b) iCoM and (c) phase of transmission reconstructed by full-field ptychography using 128 x 128 diffraction patterns, covering a large field of view of 2.5 x 2.5 $nm^2$. Both ADF and iCoM suffer from stripe artifacts and large contrast variations. In ptychography reconstruction, some of atoms are distorted and blurred, but the overall resolution is improved significantly.